\documentclass[11pt,twoside]{article}  % Leave intact
\usepackage{asp2006}
\usepackage{adassconf}

\begin{document}   % Leave intact

%-----------------------------------------------------------------------
%			    Paper ID Code
%-----------------------------------------------------------------------
% Enter the proper paper identification code.  The ID code for your paper 
% is the session number associated with your presentation as published 
% in the official conference proceedings.  You can find this number by 
% locating your abstract in the printed proceedings that you received 
% at the meeting, or on-line at the conference web site.
%
% This identifier will not appear in your paper; however, it allows different
% papers in the proceedings to cross-reference each other.  Note that
% you should only have one \paperID, and it should not include a
% trailing period.
%
% EXAMPLE: \paperID{O4.1}
% EXAMPLE: \paperID{P2.7}

\paperID{7B.3}

%-----------------------------------------------------------------------
%		            Paper Title 
%-----------------------------------------------------------------------
% Enter the title of the paper.
%
% EXAMPLE: \title{A Breakthrough in Astronomical Software Development}

\title{GPUs for data processing in the MWA}

%-----------------------------------------------------------------------
%          Short Title & Author list for page headers
%-----------------------------------------------------------------------
% Please supply the author list and the title (abbreviated if necessary) as 
% arguments to \markboth.
%
% The author last names for the page header must appear in one of 
% these formats:
%
% EXAMPLES:
%     LASTNAME
%     LASTNAME1 and LASTNAME2
%     LASTNAME1, LASTNAME2, and LASTNAME3
%     LASTNAME et al.
%
% Use the "et al." form in the case of four or more authors.
%
% If the title is too long to fit in the header, shorten it: 
%
% EXAMPLE: change
%    Rapid Development for Distributed Computing, with Implications for the Virtual Observatory
% to:
%    Rapid Development for Distributed Computing

\markboth{Ord et al. }{GPUs and the MWA}

%-----------------------------------------------------------------------
%		          Authors of Paper
%-----------------------------------------------------------------------
% Enter the authors followed by their affiliations.  The \author and
% \affil commands may appear multiple times as necessary.  List each
% author by giving the first name or initials first followed by the
% last name. Do not include street addresses and postal codes, but 
% do include the country name or abbreviation. 
%
% If the list of authors is lengthy and there are several institutional 
% affiliations, you can save space by using the \altaffilmark and \altaffiltext 
% commands in place of the \affil command.
%
% EXAMPLE: 
%      \author{Raymond Plante, Doug Roberts, 
%                  R.\ M.\ Crutcher\altaffilmark{1}}
%      \affil{National Center for Supercomputing Applications, 
%                 University of Illinois, Urbana, IL, USA}
%      \author{Tom Troland}
%      \affil{University of Kentucky, Lexington, KY, USA}
%
%      \altaffiltext{1}{Astronomy Department, UIUC}
%
% In this example, the first three authors, "Plante", "Roberts", and
% "Crutcher" are affiliated with "NCSA".  "Crutcher" has an alternate 
% affiliation with the "Astronomy Department".  The fourth author,
% "Troland", is affiliated with "University of Kentucky"

\author{S.\ Ord,  L.\ Greenhill, R. \ Wayth, D.\ Mitchell}
\affil{Harvard-Smithsonian Center for Astrophysics, Cambridge, MA, USA}

\author{K.\ Dale, H,\ Pfister, R.\ G.\ Edgar}
\affil{Harvard University, Cambridge, MA, USA}

%-----------------------------------------------------------------------
%			 Contact Information
%-----------------------------------------------------------------------
% This information will not appear in the paper but will be used by
% the editors in case you need to be contacted concerning your
% submission.  Enter your name as the contact along with your email
% address.
% 
% EXAMPLE:  \contact{Dennis Crabtree}
%           \email{crabtree@cfht.hawaii.edu}

\contact{Stephen Ord}
\email{sord@cfa.harvard.edu}

%-----------------------------------------------------------------------
%		      Author Index Specification
%-----------------------------------------------------------------------
% Specify how each author name should appear in the author index.  The 
% \paindex{ } should be used to indicate the primary author, and the
% \aindex for all other co-authors.  You MUST use the following
% syntax: 
%
% SYNTAX:  \aindex{Lastname, F.~M.}
% 
% where F is the first initial and M is the second initial (if used). Please 
% ensure that there are no extraneous spaces anywhere within the command 
% argument. This guarantees that authors that appear in multiple papers
% will appear only once in the author index. Authors must be listed in the order
% of the \paindex and \aindex commmands.
%
% EXAMPLE: \paindex{Crabtree, D.}
%          \aindex{Manset, N.}        
%          \aindex{Veillet, C.}        

\paindex{Ord, S.}
\aindex{Wayth, R.}     % Remove this line if there is only one author
\aindex{Greenhill, L.}     % Remove this line if there is only one author
\aindex{Mitchell, D.}     % Remove this line if there is only one author
\aindex{Dale, K.}
\aindex{Pfister, H}
\aindex{Edgar, R}

%-----------------------------------------------------------------------
%			Subject Index keywords
%-----------------------------------------------------------------------
% Enter up to 6 keywords that are relevant to the topic of your paper.  These 
% will NOT be printed as part of your paper; however, they will guide the creation 
% of the subject index for the proceedings.  Please use entries from the
% standard list where possible, which can be found in the index for the 
% ADASS XVI proceedings. Separate topics from sub-topics with an exclamation 
% point (!). 
%
% EXAMPLE:  \keywords{astronomy!radio, computing!grid, data management!workflows, 
%     instrumentation!control}


%-----------------------------------------------------------------------
%			       Abstract
%-----------------------------------------------------------------------
% Type abstract in the space below.  Consult the User Guide and Latex
% Information file for a list of supported macros (e.g. for typesetting 
% special symbols). Do not leave a blank line between \begin{abstract} 
% and the start of your text.

\begin{abstract}          % Leave intact
% Place the text of your abstract here - NO BLANK LINES
The MWA is a next-generation radio interferometer under construction in remote Western Australia. The data rate from the correlator makes storing the raw data infeasible, so the data must be processed in real-time. The  processing task is of order ~10 TFLOPs$^{-1}$. The remote location of the MWA limits the power that can be allocated to computing.
We describe the design and implementation of elements of the MWA real-time data processing system which leverage the computing abilities of modern graphics processing units (GPUs). The matrix algebra and texture mapping capabilities of GPUs are well suited to the majority of tasks involved in real-time calibration and imaging. Considerable performance advantages over a conventional CPU-based reference implementation are obtained. 
\end{abstract}

%-----------------------------------------------------------------------
%			      Main Body
%-----------------------------------------------------------------------
% Place the text for the main body of the paper here.  You should use
% the \section command to label the various sections; use of
% \subsection is optional.  Significant words in section titles should
% be capitalized.  Sections and subsections will be numbered
% automatically. 
%
% EXAMPLE:  \section{Introduction}
%           ...
%           \subsection{Our View of the World}
%           ...
%           \section{A New Approach}
%
% It is recommended that you look at the sample paper sample2.tex
% for examples of formatting references, footnotes, figures, equations, 
% html links, lists, and other features.  

\section{Introduction}

The Murchison Wide-Field Array (Lonsdale et al. 2007) is a 512 element, low frequency, radio interferometer, currently under construction in Western Australia. The instrument has a number of ambitious science goals, grouped under four main science packages, the detection of the Epoch of Re-Ionization; solar, heliospheric and ionospheric science; a systematic survey for radio transients; and the  Galactic and extra-galactic science package, which is a large umbrella  organization of science interests including the interstellar medium, large area surveys, pulsars and the Galactic magnetic field. The instrument is also novel in that the intention is to process and calibrate all observations in real--time, as the raw data rate is too large to capture and process offline.

\section{The Real--Time System}

An FPGA\footnote{ Field Programmable Gate Array} correlator will cross--correlate the signals received by the 512 antennas, its output being the correlations of 130,000 baseline pairs, each consisting of 768 frequency channels with 4 polarizations. This data stream is then integrated, calibrated and imaged by the real--time system (RTS).

Calibration in this sense refers to calculating the complex gain of each antenna element and the time dependent vector field that describes refractive source position shifts due to the ionosphere.   This process involves the observation of catalogue radio sources and the performance of a large linear least squares minimization in order to obtain the best estimation of the antenna gains and ionospheric offsets (Mitchell et al. 2008).  
Imaging refers to the construction, from the input data and calibration information, via a Fourier transform, of four images in the Stokes parameters, I, Q, U and V in a form that can be readily aggregated in frequency and time. 

\subsection{RTS Tasks}

The calibration tasks are applied to a data-set which can be thought of as a Fourier transform of the sky brightness distribution. These data are manipulated to measure and remove the contribution from catalogue radio sources in decreasing order of predicted flux. Each measurement being used to constrain the complex gains of the antenna elements, this is an iterative process. The data-set then undergoes a convolution that interpolates the samples onto a regular grid to permit the operation of the Fast Fourier Transform (FFT).

In the case of the MWA the subsequent FFT operation results in images that are measurements of instrumental polarization in a coordinate system that is a slant-orthographic projection of the celestial sphere. Such images cannot be directly combined (e.g. co-added) as both the projection and the instrumental polarization change as a function of time (Cornwell and Perley 1992). The RTS takes the novel approach of transforming these wide-field images into a integrable representations of the polarization state of the sky by resampling the images onto an all-sky pixelisation. Firstly the measured instrumental polarization is converted into a set of Stokes parameters by  the application of a 4x4 matrix transformation. This matrix is a function of time and position on the sky relative to the instrument and must be calculated by the RTS for each pixel. The slant-orthographic projection is converted into HEALPIX (G\'{o}rski et al 2005) via a flux redistribution algorithm requiring the calculation of input and output polygonal overlaps. The pixelisation can then be imaged via the HPX projection (Calabretta and Roukema 2007) with no further interpolation. 

Simulations and code development has indicated that the compute budget for this pipeline is approximately 10 TFLOPs$^{-1}$. The imaging tasks are by far the most computationally intensive operations. This compute capability will be provided, on site, by the real--time computer (RTC) 

\section{The RTC}

A 3.2 GHz Harpertown CPU from Intel can provide approximately 100 GFLOPs$^{-1}$. Which indicates that 100 are required to meet the RTC processing requirements. With 100 compute-nodes and assuming a conservative 300W per compute-node of power consumption results in an RTC power requirement of 30kW, not including active cooling. The RTC has to operate in the desert under very strict power limitations, the current power budget being 20kW, which indicates that the FLOP/Watt ratio of even the most recent CPU is not sufficient to perform the task. In contrast a single 2008 NVIDIA GT200 series GPU can provide 800~GFLOPs$^{-1}$. Even if the power consumption per node is raised to 600~W, 32 GPU nodes can be powered by this budget  providing 20~TFLOPs$^{-1}$ of theoretical performance.

\subsection{The GPU}

One of the fundamental problems in High Performance Computing (HPC) is that RAM access speeds have not kept pace with improvements in CPU performance. A modern CPU can operate on two floating point numbers per clock cycle, but fetching those two numbers from memory takes hundreds of clock cycles. This leads to what has been termed ``data starvation''. CPU development has attempted to overcome this problem using innovative technologies including; large caches, superscalar architecture and branch prediction.

GPU developers have taken another approach to resolving data starvation, they have devoted GPU transistors to extra execution units, each working on a single thread of execution. These threads perform the same operation on different data elements as per the SIMD (single instruction multiple data) paradigm. There are also many more threads queued for execution than are running at any one time. The goal is that whenever a thread is waiting on memory access it gets swapped out for another which has data ready for use. GPU hardware handles all the thread scheduling transparently and a modern GPU can have 100s more execution units than a modern CPU. The difficulty in utilizing this compute capability lies in providing the GPU with enough threads to hide the memory latency.

\subsection{Application of GPUs to the RTS}

Applications that benefit most from a GPU implementation are ideally massively parallel, high arithmetic intensity operations. As moving data from the host memory to device memory is a time-consuming operation, considerable effort was directed at ensuring the entire RTS pipeline could be implemented on the GPU. Even though some operations are not optimally suited, the benefit in avoiding unnecessary host to device memory transfers was significant. 

The GPU code differs from the CPU code due to the particular requirements of GPU programming. GPUs are sensitive to memory access patterns, as a result the porting involved a careful determination of optimal memory structures and even some algorithmic changes.  In addition some operations are simply inadvisable on the GPU, for example the NVIDIA CUDA API lacks an atomic floating point accumulate operation, as a result ``scatter'' type algorithms, where multiple threads may write to the same memory location are not advisable and those that exist in the CPU code have been replaced by ``gather'' type algorithms, where a thread collects the values for a particular memory location. Redundant on-device memory operations were minimized by ensuring that thread memory requests were coalesced by the GPU; if threads access consecutive memory locations, the GPU performs all the memory access operations simultaneously. GPU threads are also very ``light-weight'', and fine grained parallelism can therefore  produce remarkable performance benefits, the hardware thread manager is so efficient that it was worth considering threads that simply add two numbers together.

The majority of the RTS pipeline has been ported to the GPU and the comparative timings of those elements are presented in Table 1. The image resampling step is yet to be implemented, but this multi-step pipeline already successfully demonstrates the feasibility of a GPU based calibration and imaging system.  Experiments indicate a factor of ten improvement over the CPU application. 

\begin{deluxetable}{ccc}
\scriptsize
\tablecaption{Relative Performance of CPU and GPU Tasks. \label{GPU}}
\tablehead{
\colhead{Task} & \colhead{CPU$^\mathrm{a}$ (ms)} & \colhead{GPU$^\mathrm{b}$(ms)} } 
\startdata
CML & 250 & 22 \\
Gridding & 480  & 36 \\
FFT & 326 & 40   \\
Convolution Correction & 22.6 & 1.4 \\
Stokes Conversion & 42.4 & 4 \\
\hline
Total & 1121 & 103.4   \\
\enddata
% Text for table footnotes must follow the tabular environment but must
% be inside the table environment.  Note that it is OK to put \ref's
% in \tablenotetext.
 
%\tablenotetext{a}{Sample footnote for Table~\ref{O4.1-tbl-1}}
\tablecomments{ (a)  INTEL Core2 Quad 2.66GHz (Q9450), ASUS P5E3 Deluxe Motherboard, 4 GiB DDR3 RAM. (b) NVIDIA C1060}
\end{deluxetable}

\section{Summary}

GPUs can enable science that otherwise would be impossible due to monetary or power constraints. We have presented a general overview of a complex GPU application to calibrate and image data from the MWA, which has been ported from a CPU-based application.  Although not all elements of the RTS have  been implemented on the GPU a calibration and imaging pipeline has been demonstrated to work efficiently, and provides considerable performance improvements over a CPU implementation.

\end{document}